\documentstyle{article}[12pt]

\def\a{\alpha}
\def\b{\beta}
\def\d{\delta}
\def\D{\Delta}

\def\g{\gamma}
\def\G{\Gamma}
\def\e{\epsilon}

\def\k{\kappa}
\def\l{\lambda}

\def\m{\mu}
\def\n{\nu}
\def\s{\sigma}

\def\z{\zeta}

\renewcommand{\@}[1]{\sqrt{#1}}

\renewcommand{\le}[1]{\label{#1}\end{eqnarray}}
\newcommand{\be}{\begin{equation}}
\newcommand{\ee}{\end{equation}}
\newcommand{\bea}{\begin{eqnarray}}
\newcommand{\eea}{\end{eqnarray}}

\def\ffract#1#2{\raise .35 em\hbox{$\scriptstyle#1$}\kern-.25em/
\kern-.2em\lower .22 em \hbox{$\scriptstyle#2$}}

\let\a=\alpha\let\b=\beta\let\d=\delta
\let\e=\epsilon\let\g=\gamma

\let\k=\kappa\let\l=\lambda
\let\m=\mu\let\n=\nu
\let\s=\sigma

\let\z=\zeta\let\G=\Gamma
\let\D=\Delta

\newcommand{\del}{\partial}

\begin{document}
\rm\large \null\vskip-24pt
\begin{flushright}
UCI-2005-45\\
{\tt hep-th/0512333}
\end{flushright}
\vskip0.1truecm
\begin{center}
\vskip .5truecm {\Large\bf
On the supersymmetric completion of the $R^4$ term in \\
M-theory } \vskip 1truecm {\large\bf Arvind Rajaraman\footnote{
e-mail: {\tt arajaram@uci.edu}}
}\\
\vskip .5truecm {\it Department of Physics and Astronomy,
University of California, \\
Irvine, CA 92697, USA.}
\end{center}
\vskip .3truecm
\begin{center}
{\bf \large Abstract}
\end{center}
We examine the question of finding the supersymmetric completion of
the $R^4$ term in M-theory. Using superfield methods, we present an
eight derivative action in eight dimensions that has 32 preserved
supersymmetries. We show also that this action has a hidden
eleven-dimensional Lorentz invariance. It can thus be uplifted to
give the complete set of bosonic terms in the M-theory eight
derivative action.

\setcounter{equation}{0}


\section{Introduction}

At low energies, string theory can be reduced to an effective field
theory of the massless modes. The leading  two-derivative action for
these fields is the supergravity action $S_2$. The effective action
also contains an infinite series of higher derivative terms,
suppressed by powers of the string scale $\alpha'$, and the complete
action has the form \bea S=S_2+(\alpha')^4S_8 +(\alpha')^5
S_{10}+\dots \eea where $S_n$ contains terms with $n$ derivatives.
The leading correction in type II theories is the eight-derivative
action, which contains the famous $R^4$ term
\cite{GW,Gross:1986mw}\bea S_{8;R^4}=\int d^{10}x\ t^8t^8R^4 \eea

Previous work on the eight derivative terms has produced many
important results \cite{Green:1999qt,ElevenD,Kiritsis:1997em}. Most
importantly, several nonrenormalization theorems are known which
strongly restrict the moduli dependence of the eight-derivative
terms. In particular it is known that in IIA theory, the $R^4$ term
occurs only at tree-level and one-loop. The $R^4$ action in M-theory
can then be obtained by taking the strong coupling limit of IIA
theory \cite{ElevenD}.

There are also several other terms at the eight derivative level,
which involve the other fields of the theory (in M-theory these
fields are the gravitino and three-form field $\hat C_{MNP}$). These
terms are believed to be related to the $R^4$ term by supersymmetry.
However, little is known about the detailed structure of these terms.

\vskip 1 cm
 There are  several reasons that one wishes to know the
full action at the eight-derivative level.

At the basic level, knowledge of these terms will tell us a lot more
about actions with maximal supersymmetry, which may lead to
fundamental understandings like the off-shell nature of the theory.

From a phenomenological viewpoint, there has been a lot of interest
in flux compactifications, where fluxes are turned on in the internal
manifold (see e.g. \cite{KKLT}). This can apply both to the case of
string theory on a Calabi-Yau manifold, or M-theory on a $G_2$
manifold. The potential for moduli in this background can be
efficiently computed in the low energy effective theory, and can be
used to gain information about stable compactifications at large
radius. However, one needs to know the full action including all the
field strengths. The full action may also be needed to consider the
stabilization of the brane moduli.

Another place where the full effective action is required is for
computing corrections in Anti-de-Sitter (AdS) backgrounds, for
applications to the AdS/CFT correspondence (e.g. \cite{previous}).
These can be applied to find corrections to black hole entropy, or to
correlation functions.

\vskip 1 cm

Despite these motivations, it has not been possible so far to
determine the complete eight derivative action. Several different
approaches have been tried. All of these approaches have their own
problems.

In string theory,  the action can be computed by evaluating all the
relevant string diagrams, and extracting the low energy action
\cite{GW,Gross:1986mw,Peeters:2001ub,ampl1}. In M-theory, a similar
approach can be used using superparticle vertex operators
\cite{Peeters:2005tb}. The $R^4$ term can be found in this way.
Alternatively, one can use sigma model techniques \cite{GVZ}.

Unfortunately, string diagrams contain much more information than
just the eight-derivative terms. One needs an effective way of
extracting the low energy limit without doing the entire computation.
Furthermore, once we get to five-point amplitudes and beyond, we have
to worry about extracting contributions to the amplitude involving
the exchange of massless fields, for example those coming from a
combination of the four-point eight derivative amplitude and a tree
level three-point interaction. Furthermore, the plethora of fields in
the supergravities means that many amplitudes need to be computed.
Sigma model techniques also require intense computational effort.

 Another approach is to use the high supersymmetry of the theory.
  It is believed that the eight-derivative action is completely
determined by supersymmetry alone. One can therefore attempt to
construct the action by using the Noether method to generate terms
step by step until supersymmetry is satisfied. This has been
attempted for the heterotic string action
\cite{Romans:1985xd,Bergshoeff:1989de,deRoo:1992zp,Sch}.
Unfortunately, the vast number of fields and the plethora of possible
terms make it impractical to use this method directly in
ten-dimensional supergravity, and even the eleven-dimensional case is
very difficult.

The most promising approach is to use superfield methods. If the
complete superfield can be found, then the action can be written as
an integral over one-half of superspace. This has been attempted for
the heterotic string in \cite{Nilsson:1986rh}, and discussed for the
maximally supersymmetric theories \cite{Peeters:2001qj} (see also
\cite{Deser:nt}). For the case of M-theory, this might seem hopeless,
as there cannot be  a chiral superfield in eleven dimensions (but see
\cite{Cederwall:2000ye}.

\vskip 1cm

In this paper, we shall show that the superfield approach can indeed
be used to obtain  the complete eight-derivative effective action in
M-theory. This will require us to perform one trick: the requisite
action is constructed in eight dimensions rather than eleven. That
is, an action can be constructed which has all the required
supersymmetry and  manifest eight-dimensional Lorentz invariance. We
then show that in fact there is a hidden eleven dimensional Lorentz
invariance. Thus the eleven dimensional action can be
straightforwardly found by a dimensional oxidation of this action to
eleven dimensions.

\vskip 1 cm

The reason we need to go to eight dimensions is a natural consequence
of the structure of superfield actions. The lowest component of a
chiral superfield must be a complex scalar which does not exist in
eleven dimensions.  To get a complex scalar in a chiral theory, we
need to dimensionally reduce to eight dimensions. Hence instead of
trying to find the M-theory action directly, we will try to find the
dimensional reduction of the action on a three-torus. The
dimensionally reduced action will be constructed by superfield
methods.

In fact, we do not even need to find the full action in the lower
dimensional theory. For example, the eleven dimensional term of the
form $R^2\hat{G}_4^2$ will yield, after dimensional reduction, terms
like $R^2G_4^2$ as well as $R^2H_3^2$. If we can establish the exact
form of either of these terms in eight dimensions, we can
dimensionally oxidize to reproduce the eleven dimensional term.

 The dimensional reduction of eleven-dimensional supergravity on a
three-torus produces $N=2, D=8$ supergravity and was originally
performed in \cite{Bergshoeff:2003ri}. The scalars from the reduction
of the metric are the volume of the three-torus and 5 scalars
$L_{i}^{~m}$ ($i,m=1,2,3$). There is also the scalar $\hat C_{123}$.
These 7 scalars parametrize a $SL(3,R)/SO(3,R)\times SL(2,R)/U(1)$
coset space. In addition, the theory contains three 2-form fields, 3
gauge fields, and a three-form field. The details of this reduction
are worked out in the next section.

We then build a chiral superfield for this theory. The lowest
component of this superfield is a complex scalar built out of
$C_{123}$ and the volume of the three-torus. The curvature occurs, as
expected, with a coefficient of $\theta^4$. We can therefore expect
to obtain a supersymmetric action by integrating the fourth power of
the superfield over half of superspace.

Unfortunately, this is not the case. The reason is that we also need
a supersymmetric measure; the supersymmetric analogue of the
$\sqrt{g}$ factor.  Now it is not obvious that such a measure exists,
and in fact, in the very similar situation of type IIB supergravity,
it can be shown that such a measure does not exist
\cite{Howe,deHaro:2002vk}.
 There is a similar obstruction in our case,
and hence the supersymmetric action suggested above does not exist.

The way around this for type IIB was suggested in
\cite{Rajaraman:2005up}, and we shall apply the same reasoning here.
Instead of trying to construct the full action in eight dimensions,
we shall look for a subset of the terms.

Explicitly, we only consider bosonic terms which are composed out of
the
 curvature $R$, the three-form field strengths $H_{\m\n\rho m}$,
and the scalars $L_m^{~i}$ (each of these is uncharged under this
U(1) symmetry).
Furthermore, we consider terms which are composed out of the bosonic
terms listed above, and in addition contain 
two fermions of charge
$1/2$ and $-1/2$ respectively.

We can now go ahead and fix these terms by requiring the cancellation
of the variations. This is tedious, but the superfield approach can
help us fix these terms. Our crucial claim is that the superfield
correctly enforces this cancellation; the superfield action will thus
reproduce correctly the specific subset of the terms that we have
described above. To substantiate this claim, we perform an explicit
evaluation of the variations of the action, and explicitly show that
the variations cancel (this calculation is very similar to the one
performed in \cite{deHaro:2002vk,Rajaraman:2005up}).

We can therefore use the superfield to produce an action involving
the curvature $R$, the three-form field strengths $H_{\m\n\rho m}$,
and the scalars $L_m^{~i}$. This is sufficient, as we have mentioned,
to recover the eleven dimensional action, as long as the
eight-dimensional action has the form of a dimensionally reduced
action, that is, it should have a hidden eleven dimensional Lorentz
invariance.

We must therefore confirm that our action has this hidden Lorentz
invariance. This can be done in a straightforward way, by summing
over an entire orbit of terms generated by the eleven dimensional
rotations. The resulting action then has eleven dimensional symmetry,
and 32 supercharges. It can therefore be dimensionally oxidized to
find the eleven-dimensional action.

We close with a discussion of future directions.

\section{N=2, D=8 supergravity}

$N=2, D=8$ supergravity can be obtained as a direct dimensional
reduction of  $N=1, D=11$ supergravity to eight dimensions. The
bosonic sector of the theory contains 7 scalars, 6 vectors, 3
two-form fields, one three-form and a graviton. The fermion sector
contains two gravitinos and four fermions. We  perform the explicit
dimensional reduction, following \cite{Bergshoeff:2003ri}.

We denote the 11D fields by $ \left\{\hat{e}_{\hat{\mu}}{}^{\hat{a}},
\hat{C}_{\hat{\mu}\hat{\nu}\hat{\rho}}, \hat{\psi}_{\hat{\mu}}
\right\} $ where hatted indices run from 0 to 10. Space-time indices
are denoted $\hat{\mu}$ while tangent-space indices are denoted
$\hat{a}$.

The eleven dimensional supersymmetry variations are taken to be \bea
\d \hat e_{\hat \mu}^{~\hat a}=-{i\over 2}\bar{\e}\G^{\hat a}\hat
\psi_{\hat \mu}~~~~~~~~~~~~~~~
\\
\d \hat C_{\hat \mu\hat \nu\hat \rho}={3\over 2}\bar{\e}\G_{[\hat
\mu\hat \nu}\hat \psi_{\hat \rho]}~~~~~~~~~~~~~~
\\
\d\hat \psi_{\hat \mu}=2\hat D_{\hat \mu}\e+{i\over 144}(\G_{\hat
\mu}^{~\hat a \hat b\hat c\hat d}-8\d_{\hat \mu}^{~\hat a}\g^{\hat
b\hat c\hat d})\hat F_{\hat a\hat b\hat c\hat d} \eea

We split the coordinates $x^{\hat{\mu}} = (x^\mu, z^m)$ with
$\mu=(0,1,\ldots,7)$ and $m=(1,2,3)$. Correspondingly, we split the
indices $\hat \mu=(\mu,m), {\hat a}=({a,i})$ where $\mu, a$ run from
0 to 7, and $m,i$ run over 1, 2, 3. The bosonic fields are reduced
via the ansatz
\begin{equation}\label{Vielbein}
  \hat{e}_{\hat{\mu}}{}^{\hat{a}}   =
\left(
\begin{array}{cr}
e^{-\frac{1}{6}\varphi} e_{\mu}{}^{a} &
e^{\frac{1}{3}\varphi} L_{m}{}^{i}A^{m}{}_{\mu} \\
&\\
0             &
e^{\frac{1}{3}\varphi}L_{m}{}^{i}~~~   \\
\end{array}
\right) \,
\end{equation}
and
\begin{equation}\label{ansatzC}
\hat{C}_{abc} =  e^{\frac{1}{2}\varphi}\, C_{abc}\, , \hspace{.3cm}
\hat{C}_{abi} =  L_{i}{}^{m}B_{ab\,m}\, , \hspace{.3cm} \hat{C}_{aij}
=
 e^{-\frac{1}{2}\varphi}\, L_{i}{}^{m} L_{j}{}^{n}\,
V_{a\, mn} \, , \hspace{.3cm} \hat{C}_{ijk}=
e^{-\varphi}\epsilon_{ijk} \ell \, .
\end{equation}
The fermions are reduced by the ansatz
\begin{equation}
 \begin{array}{rcl}
{\hat \psi}_{\hat a} & = e^{\varphi/12}\left( \psi_{a} - \frac{1}{6}
\Gamma_{a} \Gamma^{i} \lambda_i \right ) \,, \qquad {\hat\psi}_{i} =
e^{\varphi/12} \lambda_i \,, \qquad \hat \epsilon = e^{-\varphi/12}
\epsilon \, .
  \end{array}
\end{equation}
We also define
\begin{equation}
  {\mathcal M}_{mn} = - L_m{}^i L_n{}^j \eta_{ij} \, ,
\end{equation}
where $\eta_{ij} = -{I}_3$ is the internal flat metric.

The dimensional reduction of the eleven dimensional field strength
$\hat G$ leads to the eight-dimensional field strengths \bea G_{\mu
\nu \rho \lambda}= 4\partial_{[\mu} C_{\nu \rho \lambda
]}+6F^{m}{}_{[\mu \nu}B_{\rho \lambda]\,m}\,,
\nonumber \\
G_{\mu \nu \rho i}=L_i^{~m}G_{\mu \nu \rho m}=
L_i^{~m}(3{\cal D}_{[\mu} B_{\nu \rho ]\,m} + 3 F^{n}{}_{[\mu \nu} V_{\rho]\,mn} ) \\
G_{\mu \nu ij}=L_i^{~m}L_j^{~n}G_{\mu \nu mn}=
L_i^{~m}L_j^{~n}(2{\cal D}_{[\mu} V_{\nu]\,mn} + \ell \epsilon_{mnp}
F^p{}_{\mu \nu})
\nonumber \\
G_{\mu ijk}=L_i^{~m}L_j^{~n}L_k^{~p}G_{\mu mnp}=
L_i^{~m}L_j^{~n}L_k^{~p}\epsilon_{mnp}\partial_\mu \ell \nonumber
\eea where the field strength of the gauge field is given by
\begin{equation}
F^m{}_{\mu \nu }  =  2\partial_{[\mu} A^{m}{}_{\nu]}
\end{equation}

The supersymmetry transformation rules in eight dimensions are \bea
\delta e_{\mu}{}^{a} = -\frac{i}{2}\overline{\epsilon} \Gamma^{a}
{\psi}_{\mu} ~~~~~~
~~~~~~~~~~~~~~~~~~~~~~~~~~  \\
\delta \psi_\mu  =  2 \partial_\mu \epsilon
 - \frac{1}{2} { \omega}_\mu^{~ab}\G_{ab} \epsilon
  +\frac{1}{2}L_{i}{}^{m} {\cal D}_{\mu} L_{mj}\Gamma^{ij} \epsilon
   +\frac{i}{96}e^{\varphi/2}(\Gamma_{\mu}^{\ \nu \rho \delta
\epsilon} -4 \delta^{\ \nu}_{\mu} \Gamma^{\rho \delta \epsilon})
G_{\nu \rho \delta \epsilon} \epsilon\nonumber \,
\\
-\frac{i}{12}e^{-\varphi}\Gamma^{ijk}
G_{\mu ijk} \epsilon
 +\frac{1}{24} e^{\varphi/2}\Gamma^iL_{i}^{\ m}
 ( \Gamma_{\mu}^{\
\nu \rho}-10\delta_\mu^{\ \nu}\Gamma^\rho )
 F_{m\nu \rho}\epsilon\nonumber
\\
+ \frac{i}{36}\Gamma^i (\Gamma_{\mu}^{\ \nu \rho \delta} -6 \delta^{\
\nu}_\mu \Gamma^{\rho \delta})G_{\nu \rho \delta i}\epsilon
 +\frac{i}{48}e^{-\varphi/2}\Gamma^{ij} 
(\Gamma_\mu^{\ \nu \rho}-10\delta_\mu^{\ \nu}\Gamma^\rho)
G_{\nu \rho ij} \epsilon   \\
\delta \psi_i  =
 \frac{1}{2}L_i^{\ m} L^{jn}{ {\cal D}}{\cal M}_{mn}
 \Gamma_j \epsilon -\frac{1}{3} \G^\mu { \partial}_\mu \varphi
\Gamma_i \epsilon +\frac{i}{24}e^{-\varphi/2}\Gamma^j 
(3\delta_i^{\ k}-\Gamma_{i}^{\ k})\G^{\mu\nu}{ G}_{\mu\nu jk}
\epsilon\nonumber
\\
 - \frac{1}{4}e^{\varphi/2}L_i^{\ m}{\cal M}_{mn}\G^{\mu\nu}{ F}_{\mu\nu}^n \epsilon
+ \frac{i}{144}e^{\varphi/2}\Gamma_i\G^{\mu\nu\rho\delta} {
G}_{\mu\nu\rho\delta}\epsilon
+\frac{i}{6}e^{-\varphi}\Gamma^{jk} 
\G^{\mu}{G}_{\mu ijk} \epsilon\nonumber\\
+\frac{i}{36}(2\delta_i^{\ j}-\Gamma_{i}^{\ j})
\G^{\mu\nu\rho}{ G}_{\mu\nu\rho j} \epsilon
 \\
\delta A^m{}_{\mu} =  -\frac{i}{2}e^{-\varphi/2} L_{i}^{\ m}
\overline{\epsilon}  (
   \Gamma^{i} \psi_{\mu} +{\Gamma}_{\mu}
(\eta^{ij}+ \frac{1}{6}\Gamma^{i}\Gamma^{j})\lambda_j ) ~~~~~~~~~~~~~~~ \\
\delta V_{\mu\, mn} =  \varepsilon_{mnp} [-\frac{i}{2}e^{\varphi/2}
L_i^{\ p} \bar \epsilon \G^9( \Gamma^i \psi_\mu +\Gamma_\mu
(\eta^{ij}-\frac{5}{6}\Gamma^i \Gamma^j) \lambda_j )-
\ell\, \delta A^p{}_\mu ]  ~~~~  \\
\delta B_{\mu \nu\, m} =  L_m^{\ \ i} \bar \epsilon (\Gamma_{i[\mu}
\psi_{\nu ]} +\frac{1}{6} \Gamma_{\mu \nu}
(3\delta_i^{\ j}-\Gamma_i \Gamma^j)\lambda_j ) -2\, \delta A^n{}_{[\mu} V_{\nu]\, mn} ~~~~~~~~ \\
\delta C_{\mu \nu \rho} =  \frac{3}{2}e^{-\varphi/2} \bar \epsilon
\Gamma_{[\mu \nu}( \psi_{\rho]} -\frac{1}{6}\Gamma_{\rho]} \Gamma^i
\lambda_i )
-3 \delta A^m{}_{[\mu} B_{ \nu \rho]\,m} ~~~~~~~~~~~~~ \\
\delta \varphi
   =  -\frac{i}{2} \overline{\epsilon} \Gamma^{i} \lambda_i ~~~~~~~~~~~~~
   ~~~~~~~~~~~~~~~~~~~~~~~~~~~~~~~~~~~ \\
\delta \ell =  -\frac{i}{2}e^{\varphi} \bar \epsilon \G^9\Gamma^i
\lambda_i ~~~~~~~~~~~~~~~~~~~~~~~~~~~~~~~~~~~~~~~~~~\eea

Here we have defined $\G^9=i\G^{123}$.

This theory has a manifest $SL(3,R)$ acting on the compactification
three-torus. There is also a $SL(2,R)$ symmetry, which corresponds to
the electric-magnetic duality of 11-dimensional supergravity. We will
now rewrite the fields to make this more manifest.

To represent the $SL(2, R)$ symmetry linearly on the scalars, we must
introduce an extra compensating scalar $\phi$. The scalars are
organized into a $SL(2,R)$ matrix \bea V= {1\over
\sqrt{2i}}\left(\begin{array}{cc} u
   & u^*\\ v& v^*\end{array}\right)\equiv {1\over
\sqrt{2i\tau_2}}\left(\begin{array}{cc} \bar\tau e^{-i\phi}
   & \tau e^{i\phi}\\ e^{-i\phi}& e^{i\phi}\end{array}\right)
   \eea
Here $\tau=l+ie^\phi$ parametrizes the upper half plane. There is now
a local U(1) action that acts as a shift on the angular variable
$\phi$,
   and which can be used to set $\phi=0$.

We will define $ V_{\mu\, mn} =\e_{mnp}W_{~\mu}^{p}$.
  Then the potentials $(A,W)$ form a $SL(2,R)$ doublet. They can be
  organized into $SL(2,R)$ invariant fields defined by
   \bea (a^m_{~~\mu},(a^{m}_{~~\mu})^*)=\sqrt{2i}( A^m{}_{\mu},
W_{~~\mu}^{m})V
 \eea
 with the corresponding field strengths $f_{\mu\nu}=2\del_{[\mu}
 A^{~m}_{\nu]}$.

  Similarly, the four-form field strength $G_{mnpq}$ and its dual four-form
  $\tilde{G}_{mnpq}$ can be organized into $SL(2,R)$ invariant field
  strengths by
  \bea
({F}_{\mu \nu \rho \lambda},{F}^*_{\mu \nu \rho \lambda})=
\sqrt{2i}(G_{\mu \nu \rho \lambda},\tilde{G}_{\mu \nu \rho \lambda})V
  \eea

  In the fermion sector, we define
  \bea \zeta={(1+i\G^9)\over 2}\e\qquad
\tilde\psi_\mu={(1+i\G^9)\over 2}\psi_\mu\qquad \l={(1+i\G^9)\over
2}\G^i\psi_i
\\
\chi_m={(1-i\G^9)\over 2}(\psi_m-{1\over 3}\g_m\g^i\psi_i
)~~~~~~~~~~~~~~~~ \nonumber\eea

Now we can rewrite the supersymmetry transformation laws

\bea \delta e_{\mu}{}^{a} = -\frac{i}{2}\left(\zeta\G^0 \Gamma^{a}
\tilde{\psi}^*_{\mu}+\zeta^*\G^0 \Gamma^{a} \tilde{\psi}_{\mu}
\right)~~~~~~ ~~~~~~~~~~~~~~~~~~~
\\
 \delta \tilde\psi_\mu  =  2 \nabla_\mu \zeta
  +\frac{1}{2}L_{i}{}^{m} {\cal D}_{\mu} L_{mj}\Gamma^{ij} \zeta \,
 +\frac{i}{24} \Gamma^iL_{i}^{\ m} ( \Gamma_{\mu}^{\
\nu \rho}-10\delta_\mu^{\ \nu}\Gamma^\rho )
 f^*_{m\nu \rho}\zeta^*\nonumber
\\
 -\frac{1}{192}\G^{\m\n\rho\s} F_{\m\n\rho\s}^*\g_\mu
\zeta^* + \frac{i}{36}\Gamma^iL_i^{\ m}(\Gamma_{\mu}^{\ \nu \rho
\delta} -6 \delta^{\ \nu}_\mu \Gamma^{\rho \delta})G_{\nu \rho \delta
m}\zeta
\\
 \delta\l
=\g^ap_a\zeta^*
 - \frac{i}{4}\G^i L_i^{\ m}\G^{\m\n}{ f}_{\m\n m} \zeta
+ \frac{1}{96} \G^{\m\n\rho\s}{F}_{\m\n\rho\s}\zeta~~~~~
\\
\d\chi_i  =
 \frac{1}{2}L_i^{\ m} L^{jn} \G^\mu{\cal D}_\mu{\cal M}_{mn}
 \Gamma_j \zeta
 - \frac{i}{12}(3\d_i^{~j}-\g_i\g^j)
 L_j^{\ p}{ f}^*_{p\m\n}\g^{\m\n} \zeta^*\nonumber\\
+\frac{i}{36}(3\delta_i^{\ j}-\Gamma_{i}\G^{j})L_j^{\
m}\G^{\m\n\rho}{ G}_{\m\n\rho m} \zeta~~~~~~~~~~~~~~~~~~~~~~~~
\\
 \d a^{m}_{~~\mu}
=- L_i^{\ m}\left(  \zeta \G^0\Gamma^i \tilde\psi_\mu + \zeta\G^0
\Gamma_\mu \chi^i +\frac{1}{2} \zeta^* \G^0\Gamma_\mu \Gamma^i
\lambda\right)~~~~~~~~~~
\\
\delta B_{\mu \nu\, m} =  L_m^{\ \ i} \z\G^0 (\Gamma_{i[\mu}
\tilde\psi^*_{\nu ]} +\frac{1}{2} \Gamma_{\mu \nu} \chi^*_i ) +
L_m^{\ \ i} \z^*\G^0 (\Gamma_{i[\mu} \tilde\psi_{\nu ]} +\frac{1}{2}
\Gamma_{\mu \nu}
\chi_i ) \nonumber\\
+i \e_{mnp}(a^n_{~~\mu}\d a^{*p}_{~~\nu}-a^{*n}_{~~\mu}\d
a^{p}_{~~\nu}) ~~~~~~~~
\\
\d u=-{1\over 2iv}\zeta\G^0 \lambda~~~~~~~~~~~~~~~~~~~~~~~~~~~~
 \eea

The various field strengths are now charged under the local U(1)
symmetry. The scalars $u,v$ have charge 1. The field strengths
$f_{mn}$ and $\hat{F}$ have charge 1, while the two-form field
$B_{\m\n}$ and the curvature have zero charge. The gravitino
$\tilde{\psi}_\mu$ and the fermion $\chi_m$ have charge $1/2$, while
the fermion $\l$ has a charge $3/2$.


\section{The linearized superfield}

We now start the superfield analysis of the theory.

The superspace coordinates are $(x^\mu,\theta^\a,\theta^{*\bar\a})$,
where $\theta^{*\bar\a}=(\theta^\a)^*$, and $\theta$ is a 16
component Weyl spinor satisfying $(1+i\G^9)\theta=0$.
 The supersymmetric derivatives are defined as
\bea D_\a^j = {\del \over {\del \theta^\a}} + i \theta^{*\bar\a }
\G^\mu_{\a\bar\a} \del_\mu \qquad \bar{D}_{\bar\a} = {\del \over
{\del \theta^{*\bar\a}}} + i \theta^{\a } \G^\mu_{\a\bar\a} \del_\mu
\eea

The chiral superfield satisfies $\bar{D}\Phi=0$, and has for its
lowest component the scalar $u$. The rest of the superfield can be
determined from the basic equation for the supersymmetry variation of
any superfield V \bea\label{supvar} \d_\z V=\z^\a D_\a V-\z^{*\bar\a}
\bar{D}_{\bar\a}V \eea Repeatedly applying this equation, we find the
components of the chiral superfield.

We find for the first few components at the linearized level \bea
\Phi|_{\theta=0}=u~~~~~~~~~~~~~~~~~~~~~~~~~~~~~~~~~~~~~
\\
D_\a\Phi|_{\theta=0}={1\over
2iv}(\G^0\l)_\a~~~~~~~~~~~~~~~~~~~~~~~~~~~~~~~~~
\\
D_{[\a} D_{\b]} \Phi|_{\theta=0}= -{1\over 2iv}\left( \frac{i}{4}\G^i
L_i^{\ m}\G^{\m\n}\G^0{ f}_{\m\n m}  + \frac{1}{96}
\G^{\m\n\rho\s}\G^0{F}_{\m\n\rho\s}\right)_{\b\a}
 \\
D_{[\a} D_{\b}D_{\g]} \Phi|_{\theta=0} = {1\over 8v} \left[(\G^i
\G^{\m\n} \G^0)_{\g\b} \left( (\G^0\Gamma^i)_{\a\d}
\tilde\psi_{\mu\nu}^\d - 2 (\G^0\Gamma_{[\mu})_{\a\d}
\del_{\nu]}\chi^{i\d}\right)\right.\nonumber
\\
\left.+\frac{1}{4}\left( \G^{\m\n\rho\l}\G^0\right)_{\g\b}
(\G^0\Gamma_{[\mu \nu})_{\a\d}\tilde\psi^\d_{\l\rho]} \right]\eea

We now work out the terms in the next order of the superfield which
are proportional to the curvature. These terms are
 \bea
 D_{[\a} D_{\b}D_{\g}D_{\d]} \Phi|_{\theta=0} = -{1\over 16v} \left[(\G^i
\G^{\l\rho} \G^0)_{\d\g} (\G^0\Gamma^i\G_{\s\tau})_{\b\a}~~~~~~~~~
 \right.\nonumber\\ \left.
 +\frac{1}{4}\left( \G^{\m\n\rho\l}\G^0\right)_{\d\g}
(\G^0\Gamma_{\mu \nu\s\tau})_{\b\a}\right]R_{\l\rho}^{~~\s\tau} \eea

The terms in the superfield multiplied by $\theta^4$  all have two
derivatives. Thus if we integrate $\Phi^4$ over half of superspace,
we will produce an eight derivative action \bea S_8=\int d^{10}x
d^{16}\theta \Phi^4
 \eea which will have linearized
supersymmetry.

When we include the moduli, we can have a moduli-dependent
coefficient multiplying this action; this coefficient is not itself
determined by linearized supersymmetry. Explicit computations in
string theory show that the above action must be multiplied by the
function $ln(\eta(u))$. Hence the action would have the form (up to
an overall constant)  \bea  S_8=\ln(\eta(u))\int d^{10}x d^{16}\theta
\Phi^4
 \eea

\section{The Nonlinear Action}

We can now try and extend this to the nonlinear case.

When we try to go beyond the quartic action, we will need the full
nonlinear superfield. In addition we need a supersymmetric measure;
the supersymmetric analogue of the $\sqrt{g}$ factor. The suggested
form of the eight-derivative action is then \bea \label{act1}
S_8=\int d^{10}x \int d^{16}\theta \Delta
\Phi^4~~~~~~~~~~~~~~~~~~~~~~~~~~~~~~~~~~~~~~~~~~~~~~~~~~ \nonumber
\\
= \int d^{10} x\, \e^{\a_1..\a_{16}} \sum_{n=0}^{16} {1 \over n! (16
-n)!} D_{\a_1} ..D_{\a_{n}} \D|\, D_{\a_{n+1}}...D_{\a_{16}} W|\eea
where $W=\Phi^4$, and $\Delta$ is by definition a superfield whose
lowest component is \bea \label{Delta0}\Delta|_{\theta=0}=\sqrt{g}
\eea $\Delta$ is to be constructed order by order by requiring that
the action be supersymmetric.

Now it is not obvious that such a measure exists, and in fact, in the
very similar situation of type IIB supergravity, it can be shown that
such a measure does not exist \cite{Howe,deHaro:2002vk}. The issue is
that while we can arrange that all variations proportional to $\z$
cancel, the variations proportional to $\z^*$ will then not cancel.
There is a similar obstruction in this case, and hence the
supersymmetric action suggested above does not exist.

In \cite{Rajaraman:2005up}, it was shown that despite this problem,
there was still some nontrivial information available from the
superfield expression. In particular, a subset of the terms in the
action is correctly generated from the superfield. The same reasoning
will apply here.

To make this explicit, we now define the subset of the terms that we
will look at. \vskip 1 cm

We restrict attention to the bosonic terms which involve only the
field strengths which are uncharged under the $U(1)$, viz. the
curvature $R_{\m\n\rho\s}$, the three-form field strengths
$H_{\m\n\rho m}$, and the scalars $L_m^{~i}$. Examples of such terms
are $R^4, R^2H^4$ etc.

Now under a supersymmetry transformation, these terms produce
variations which contain one or more fermion fields; for instance,
the variation of the $R^4$ term will produce variations of the
generic form $R^3\bar{\z}D^2\psi$. This must be cancelled by the
variation of terms bilinear in fermions, for example, a term of the
form $R^2D\psi D^2\psi$. An analysis of the U(1) structure shows that
these terms must be of a particular form: they involve the uncharged
fields $R_{\m\n\rho\s}$, $H_{\m\n\rho m}$, and $L_m^{~i}$, and
 in addition they have two fermions, one of which carries a $1/2$
charge under the U(1) (i.e. $\psi_\mu$ or $\chi_a$), and one with a
$-1/2$ charge under the U(1) (i.e. $\psi^*_\mu$ or $\chi^*_a$).

  We fix these terms by requiring a cancellation of the
variations. It will suffice to consider those variations which have
at most one fermion field i.e. we ignore the cancellation of the
terms with three fermions. The cancellation of variations with one
fermion will be enough to determine the subset of bosonic terms in
the action that we are considering.

Our crucial claim is that the superfield correctly enforces this
cancellation; the superfield action (\ref{act1}) will thus reproduce
correctly the specific subset of the terms that we have described
above.

To prove this, we start by noting that the uncharged field strengths
are all found in the $\theta^4$ component of the superfield. The
fermionic terms that we are considering are all to be found in the
$\theta^3,\theta^5$ components. Thus, when we look for the bosonic
terms in the action, the 16 $\theta$ are then already saturated from
the $\Phi^4$ term. For the terms bilinear in fermions, at least 15
$\theta$ must be taken from the $\Phi^4$ term (as opposed to factors
of $\theta$ coming from $\Delta$).

Hence to construct the action, we only need the first two components
of $\Delta$, i.e. $\Delta|_{\theta=0}\equiv \sqrt{g}$ and
$D_\a\Delta|_{\theta=0}$. We do not need the other components of the
measure, as long as we are restricting ourselves to this particular
subset of terms.

To summarize, we are setting
$\del{\tau}=\l_i=a^m_\mu=C_{\m\n\rho}=0$, and we are considering
variations with at most one fermion field. We may truncate the action
to \bea \label{act2} S = \int d^{8} x\, {1
\over 16!}
\e^{\a_1..\a_{16}} \left(\sqrt{g}\, D_{\a_{1}}...D_{\a_{16}} W| +16
D_{\a_1}\D|\, D_{\a_{2}}...D_{\a_{16}} W|\right) \eea

 We now need to show
that this action is supersymmetric, and we shall do this in the next
section. This analysis will follow \cite{deHaro:2002vk} closely.

\section{Cancelling the Supersymmetry Variations}

To analyze the supersymmetry variations, we will need some facts
about the torsions. These are determined by the algebra \be [D_A, D_B
\} = - T_{A B}{}^C D_C + \frac{1}{2} R_{ABC}{}^D L_D{}^C + 2i M_{AB}
\k~, \ee

We can set some torsions and curvatures to zero because there are no
terms of the right dimension and charge. We then find that the
nonzero torsions are $T_{\a \b}{}^{\bar\g}, T_{\a \bar\b}{}^c, T_{a
\b}{}^\g, T_{a \b}{}^{\bar\g}, T_{a b}{}^\g$ and their complex
conjugates.

The curvatures are determined from the torsions by the Bianchi
identities \bea 
\sum_{(ABC)} (D_A T_{BC}{}^{D} + T_{AB}{}^{E} T_{EC}{}^D -
\hat{R}_{ABC}{}^{D}) = 0 \eea in particular \bea \label{Req}
T_{\bar\a\b}{}^{c} T_{c\g}{}^{\d}+ T_{\bar\a\g}{}^{c} T_{c\b}{}^{\d}
+ T_{\b\g}{}^{\bar{\e}} T_{\bar{\e}\bar\a}{}^{\d} -
\hat{R}_{\bar\a\b\g}{}^{\d} = 0 \eea

The torsions can be determined from the supersymmetry algebra.For
example, we have \bea D_a D_\b V-D_\b D_a V=- T_{a\b}{}^\g D_\g V
\eea

Noting that \bea D_a \equiv e_a^M D_M
 =e_a^{~m} D_m  -{1\over 2}\psi^\a_a D_\a
 +{1\over 2}\psi^{*\bar\a}_a D_{\bar\a}
 \eea
 we find that the algebra implies that
 \bea
T_{\b \bar\a}^{~~c}=-{i}(\G^0\Gamma^{c})_{\b\bar\a}
\eea 
and
 \bea T_{a\b}{}^\a=
 -\frac{1}{4}L_{i}{}^{m} {\cal D}_{\mu} L_{mj}(\Gamma^{ij})_{\a\b}
 - \frac{i}{72}(\Gamma^iL_i^{\ m}(\Gamma_{\mu}^{\ \nu \rho
\delta} -6 \delta^{\ \nu}_\mu \Gamma^{\rho \delta}))_{\a\b}G_{\nu
\rho \delta m} \eea

Now we return to the considerations of the supersymmetry variations.
Once again, we are setting $\del{\tau}=\l=a_2=0$, and we are
considering variations with at most one fermion field. We can then
set $D^{n}W|=0$ in the supersymmetry variations for all $n\leq 14$.
 We thus only need to cancel the
variations proportional to $D^{16}W$ and  $D^{15}W$. Furthermore we
can set $[D_{\a_{1}}, D_{\a_{2}}] \Delta|=0$, since it has a U(1)
charge of 1.

The variations are then \bea \d S = \int d^{10} x \left[ \d e
D^{16} W| + e (\d D^{16} W|) + e 
D_{\a} \Delta| \d D^{15,\a} W|
+e\d D_{\a} \Delta| D^{15,\a} W|
 \right]
 \eea

Consider each term separately.

For the first term, the variation of $e$ is \bea \d e
 =-{i\over
2}ee^\mu_{~a}(\z\G^0\G^a\psi_\mu^*+\z^{*}\G^0\G^a\psi_\mu) \eea

In the second term, the variation of the $D^{16} W|$ term is \be \d
D^{16} W|=\frac{1}{16!} \e^{\a_1 ...\a_{16}} (\zeta^{\a} D_{\a} -
\zeta^{*\bar\a} \bar D_{\bar\a} ) D_{\a_1} ... D_{\a_{16}} W|~. \ee
In the $\z$ terms we can antisymmetrize the $D_\a$ derivatives, and
since there are only 16 $D_\a$, this term is zero. For the $\z^*$
terms, we compute the commutator $[\bar D_{\a}, D^{16}] W|$. We find
\bea \label{d*1} \d D^{16} W|&=&- \zeta^{ *\bar\a}({1\over
2}T_{\bar{\a} \d}^{~~c} \psi_c^{\d} D^{16} W| )+ \zeta^{
*\bar\a}\left(e_c^m T_{\bar{\a} \b}^c D_m  + T_{\bar{\a} \g}^c T_{\b
c}^{\g} \right) D^{\b, 15} W| \nonumber \eea where we have used
(\ref{Req}), and dropped the torsions with $U(1)$ charge greater than
1/2.

 In the third term\be \frac{1}{15!}
\d D^{15,\a} W
= \frac{1}{15!}\e^{\a\a_2 .. \a_{16}} (\zeta^{\b} D_{\b} - \zeta^{*
\bar\b} \bar D_{\bar\b} )( D_{\a_2} .. D_{\a_{16}} W|)
 \ee
 The $\z^*$ terms are all of the form $D^nW$ with $n<15$, and can
 be ignored. So \be\frac{1}{15!}
\d D^{15,\a} W
=\zeta^{\a}D^{16} W|
 \ee

We can now calculate the total coefficient of $D^{16}W$. This is \bea
\z^\a\left(-{i\over 2}ee_\mu^a\G^0\G^a\psi_\mu^*-e
D_{\a}\Delta|\right)+\z^{*\bar\a}\left(-{i\over
2}ee_\mu^a(\G^0\G^a)_{\bar\a\b}\psi^\b_\mu
 -e{1\over 2}T_{\bar{\a} \d}^{~~c} \psi_c^{~\d}\right)
\eea

The second term cancels. From the first term, we learn that we must
take\bea D_{\b}\Delta| =-\frac{i}{2}(\G^0\Gamma^{a})_{\b\bar\a}
{\psi}_a^{*\bar\a}=\frac{1}{2}T^{a}_{\b\bar\a} {\psi}^{*\bar\a}_a\eea

The coefficient of $D^{\b, 15} W|$ in the variation is then \bea
\frac{1}{2}T^{a}_{\b\bar\a} \d{\psi}^*_{a\bar\a} 
+ \zeta^{ *\a}\left(e_c^m
T_{\bar{\a} \b}^c D_m  
+  T_{\bar{\a} \g}^c T_{\b c}^{\g}  \right) 
\nonumber \eea

When taking the variation of the gravitino, the terms proportional to
$\z$ all multiply terms with U(1) charge greater than 1/2, and can be
dropped. The terms proportional to $\z^*$ are easily shown to cancel
in the above coefficient up to a total derivative.

To summarize, we have shown here that the action (\ref{act2}) is
invariant under supersymmetry transformations, after we perform the
truncation described in the previous section.

\section{Lorentz Invariance}

Let us review what we have found so far. We have found supersymmetric
actions of the form (\ref{act2}). In these expressions, $\Phi$ is a
chiral superfield, but we only need
 the $\theta^{3,4,5}$
terms i.e. the superfield $\Phi$ can be truncated to $ \Phi\sim
\theta^3\Phi_3+\theta^4\Phi_4+\theta^5\Phi_5 $. Then (\ref{act2})
will provide a supersymmetric expression as long as $\Phi_3$ is a
linear combination of terms which have $U(1)$ charge 1/2. In our
case, there are two fermions $\tilde\psi_{\mu\nu}$ and
$\del_\mu\chi_i$ which have this charge, and the proper dimension,
and so $\Phi_3$ in general can be taken to be a linear combination of
these objects.

The correct linear combination, which we denote $\Phi_{inv}$, can be
determined by requiring the action to have eleven-dimensional Lorentz
invariance.

 Let us  suppose
 we want to extend an SO(2) invariant object to an SO(3) invariant
 object. For example, take the SO(2) invariant
 $A_xB_x+A_yB_y\equiv \sum_{i=1,2}A_iB_i$. Then the SO(3) invariant object is
 immediately found to be $\sum_{i=1,3}A_iB_i$, that is, we simply
 extend the sum over all possible indices. The same principle can
 be applied to our case.

Now in our case, the third term in the superfield contains the term
\bea \theta^\g \theta^\b \theta^\a D_{[\a} D_{\b}D_{\g]}
\Phi|_{\theta=0} = ...+\frac{i}{4} \bar\theta\G^{\m\n\rho\l}\theta
\bar\theta\Gamma_{[\mu \nu}\tilde\psi_{\l\rho]}
 \eea
Here $\m,\n...$ run from 0 to 7. To make this Lorentz invariant, we
should extend the sum over the eleven dimensional indices. For the
gamma matrices, for instance, we must add terms where  $\G^\mu$ has
been replaced with $\G^i$.

For the gravitinos, we should use the relation between the eleven
dimensional gravitino and the eight-dimensional gravitino
\begin{equation}
 \begin{array}{rcl}
{\hat \psi}_{\hat a} & = e^{\varphi/12}\left( \psi_{a} - \frac{1}{6}
\Gamma_{a} \Gamma^{i} \lambda_i \right )
  \end{array}
\end{equation}
Now we are setting $\l_i=0$ in all terms. We are also ignoring the
moduli dependence. At this level of approximation, we can write the
above term as \bea \theta^\g \theta^\b \theta^\a D_{[\a}
D_{\b}D_{\g]} \Phi|_{\theta=0} = . ...+\frac{i}{4}
\bar\theta\G^{\m\n\rho\l}\theta
\bar\theta\Gamma_{[\mu \nu}\hat\psi_{\l\rho]}
 \eea

The extension of the term to a Lorentz invariant form is now
straightforward; we thus get the Lorentz invariant object \bea
W_3={i\over 4}\left( \bar\theta\G^{\m\n\rho\l}\theta
\bar\theta\Gamma_{\mu
\nu}\hat\psi_{\l\rho}+\bar\theta\G^{ij\rho\l}\theta
\bar\theta\Gamma_{ij}\hat\psi_{\l\rho}+4\bar\theta\G^{i\n j\l}\theta
\bar\theta\Gamma_{i \nu}\hat\psi_{j\l} +\bar\theta\G^{\m\n ij}\theta
\bar\theta\Gamma_{\m \nu}\hat\psi_{ij}\right) \eea
 We have used the
fact that $\theta$ is a Weyl spinor to simplify the expression.

This is the $\theta^3$ component of the required Lorentz invariant
superfield $\Phi_{inv}$. To construct the action, we also need the
$\theta^4, \theta^5$ components of $\Phi_{inv}$. These can be found
by using the standard formula (\ref{supvar}). We shall leave the
explicit evaluation of these terms to a future paper, and here we
will summarize these terms by formally replacing $W_3$ by the
expression \bea W={i\over 4}\left( \bar\theta\G^{\m\n\rho\l}\theta
\bar\theta\Gamma_{\mu
\nu}\tilde\Psi_{\l\rho}+\bar\theta\G^{ij\rho\l}\theta
\bar\theta\Gamma_{ij}\tilde\Psi_{\l\rho}+4\bar\theta\G^{i\n
j\l}\theta \bar\theta\Gamma_{i \nu}\tilde\Psi_{j\l}
+\bar\theta\G^{\m\n ij}\theta \bar\theta\Gamma_{\m
\nu}\tilde\Psi_{ij}\right) \eea We have here defined the new
superfields $\tilde\Psi_{\l\rho}, \tilde\Psi_{j\l}, \tilde\Psi_{ij}$.
The lowest components of these superfields are respectively
$\hat\psi_{\l\rho}, \hat\psi_{j\l}, \hat\psi_{ij}$. ($W$ itself is
not a superfield; it should be thought of as the sum of the
$\theta^3,\theta^4, \theta^5$ terms of the superfield $\Phi_{inv}$.)

Including the moduli-dependent coefficient, the full action is then
(up to an overall constant)
 \bea  S_8 = \ln(\eta(u))\int d^{8}
x\, {1
\over 16!}
\sqrt{g}\e^{\a_1..\a_{16}} \left(\, D_{\a_{1}}...D_{\a_{16}} W^4|
-8{i}(\G^0\Gamma^{a})_{\a_1\bar\a} {\psi}_a^{*\bar\a}
D_{\a_{2}}...D_{\a_{16}} W^4|\right) \eea

This action has manifest $N=2$ supersymmetry in 8 dimensions (after
the truncation already described), and
 is clearly the reduction
of an action with 11-dimensional Lorentz invariance.  To find the
explicit M-theory action, we need to evaluate the suerspace
derivatives (or alternatively, perform an integration over the
superspace coordinates) and obtain the action in coordinate space.
The resulting action can be dimensionally oxidized to eleven
dimensions.


\section{Discussion}

We have found part of an action in eight dimensions which has 32
supersymmetries. This action encodes all terms in eight dimensions
involving the curvature $R_{\m\n\rho\s}$ and the three forms
$H_{\m\n\rho m}$. In future work, we will uplift this action to
obtain all the bosonic terms in the eight-derivative M-theory
effective action. It should also be possible to use our technique to
find the terms bilinear in fermions.

In addition to finding the explicit action, there are several
directions of interest to pursue.

Knowledge of the M-theory action allows us to find the one-loop type
IIA action by a dimensional reduction. It would be interesting to
develop techniques to fix the tree level part of the type IIA action
as well. Similarly, we would like to work out the action for M-theory
compactified on arbitrary tori.

More speculatively, we may be finding hints about the {\it off-shell}
superspace formulation of the theory. Little is known currently about
the off-shell superspace formulation of theories with 32
supercharges; even the auxiliary field content has not been
determined. Our results here suggest that if such a formulation
exists, it should exist in eight dimensions rather than eleven. It
may be that to obtain a manifestly supersymmetric formulation, we
have to give up manifest Lorentz invariance. It would be very
interesting to see if our results can be extended to make this
explicit; understanding the structure of the fermion bilinears will
also help in this.

\vskip 0.5 cm
 {\bf Acknowledgements}: We are grateful to
N.~Berkovits, S.~Deser, S.~de Haro, M.~Green, P.~Howe, A.~Sinkovics
and K.~Skenderis for useful comments.

The author is supported in part by NSF Grant PHY-0354993.

\end{document}